\documentclass[letter]{aa}  

\usepackage[colorlinks=true,linkcolor=blue,citecolor=blue,filecolor=blue,urlcolor=blue]{hyperref}
\usepackage{graphicx}
\usepackage{txfonts}
\begin{document}

   \title{Intermediate-mass black hole incubators}

   \subtitle{Gas accretion onto stellar black hole clusters in galactic central molecular zones}

   \author{Jaroslav Haas\inst{1}\thanks{\email{haas@sirrah.troja.mff.cuni.cz}},
           Pavel Kroupa\inst{1,2}
           \and Sergij Mazurenko\inst{3}}

   \institute{Charles University, Faculty of Mathematics and Physics, Astronomical Institute,
              V Hole\v{s}ovi\v{c}k\'ach 2, CZ-18000 Prague, Czech Republic
              \and
              Helmholtz-Institut f\"{u}r Strahlen- und Kernphysik, University of Bonn,
              Nussallee 14-16, D-53115 Bonn, Germany
              \and
              University of Bonn, Regina-Pacis-Weg 3, D-53113 Bonn, Germany
             }

% \abstract{}{}{}{}{}
% 5 {} token are mandatory
 
  \abstract
  % context heading (optional)
  % {} leave it empty if necessary  
   {The stellar dynamical evolution of massive star clusters formed during starburst periods
   leads to the segregation of $\gtrsim10^4\,M_\odot$ stellar-mass black hole sub-clusters in their
   centres. In gas-rich environments, such as galactic central molecular zones, these black hole
   clusters are likely to accrete large amounts of the gas from their surroundings, which in
   turn affects their internal dynamics.}
  % aims heading (mandatory)
   {In this Letter we estimated the corresponding accretion rate onto the black hole cluster and
   its radiative feedback. We assessed whether such an accretion flow can lead to the collapse
   of the black hole cluster into an intermediate-mass black hole.}
  % methods heading (mandatory)
   {The estimates were obtained analytically, considering the astrophysical conditions and star
   formation history reported for the central molecular zone of our Galaxy.}
  % results heading (mandatory)
   {We find that a stellar black hole cluster with mass $\gtrsim10^4\,M_\odot$ located in the
   twisted ring of molecular clouds with radius $\approx100$~pc that is observed in the
   central molecular zone of our Galaxy can accrete about the same mass in gas on a timescale
   of a few million years. We suggest that this is sufficient for its subsequent collapse into
   an intermediate-mass black hole. Based on an estimate of the dynamical friction inspiral time,
   we further argue that the locations of the intermediate-mass black hole candidates recently
   observed in the central molecular zone are compatible with their formation therein during
   the last starburst period reported to have occurred $\approx1$~Gyr ago.}
  % conclusions heading (optional), leave it empty if necessary
   {}


   \maketitle
%
%--------------------------------------------------------------------------------------------------
%
\section{Introduction}
\label{intro}
Black holes were theoretically predicted more than a century ago. Their observational
discovery in the Universe, however, had to wait many decades for the technological advances
to catch up with theory. The existence of the stellar-mass black holes
($\lesssim10^2\,M_\odot$) representing the remnants of stellar evolution
has been gradually proven beyond any reasonable doubt
\citep{Webster72,LIGO-Virgo16a,LIGO-Virgo16b,Sahu22}. The same holds for
the supermassive black holes
\citep[SMBHs; $\gtrsim10^5\,M_\odot$;][]{Eckart96,Ghez98,GRAVITY22,EHT19,EHT22} that are widely
accepted to reside in the centres of major galaxies.

The intermediate-mass black holes (IMBHs) with masses
$\approx10^{2-5}\,M_\odot$, however, have been eluding discovery so far. In an intuitive
agreement with their masses, the theoretically formulated stellar dynamical formation
scenarios for IMBHs typically place them at the centres of massive globular
clusters and dwarf galaxies \citep[see][for reviews]{Mezcua17,Greene20} or in regions near
SMBHs embedded in dense galactic nuclear star clusters
\citep[see][and references therein]{Rose22}.
A notable candidate was most recently reported in the centre of $\omega$~Centauri by
\citet{Haberle24}.

Nevertheless, quite independently of any formation scenario, five IMBH
candidates have also been reported
\citep{Maillard04,Oka16,Tsuboi17,Takekawa19a,Takekawa19b,Takekawa20} in the central molecular
zone (CMZ) of our Galaxy, a region rich in gas \citep{Morris96,Battersby20,Battersby25}
that extends to about 250~pc in radius from the central SMBH
Sgr~A$^\star$ \citep{GRAVITY22}. Observations of
the candidates typically revealed gas (or stellar) motions that
are best described by orbits around a massive and very compact source of gravitational
potential.

In this Letter we present a new formation scenario for IMBHs in such
environments that relies on agents naturally present there. In particular, the
driving effect of the IMBH formation in the presented model
is the accretion of the abundant gas onto massive stellar black hole clusters assumed to be
the remnants of massive star clusters that form during starburst periods in gas-rich regions.
%
%--------------------------------------------------------------------------------------------------
%
\vspace*{-2mm}
\section{Collapse of the black hole cluster flooded by gas to an intermediate-mass black hole}
\label{shrinkage}
An effective formation mechanism for SMBHs was revealed by \citet{Kroupa20}
that enables them to emerge soon enough after the Big Bang to explain the cosmological
observations. In their model, the SMBHs form through a two-phase collapse
of a hypermassive cluster of stellar-mass black holes at the centre of the forming galaxies.
These stellar-mass black holes are considered to be the remnants of the primordial very massive
stars born in the early metal-free Universe at the centres of collapsing pre-galactic gas
clouds. In the first phase, the stellar black holes are pushed together by the gas inflowing
from the farther parts of the forming galaxy owing to their
friction on the gas and due to the deepening of the gravitational potential well caused
by the accumulation of the gas within the cluster. Once the cluster becomes compact enough,
the second phase of the collapse driven by the emission of gravitational waves starts, thus leading
to a swift formation of the SMBH seed through mergers of the stellar
black holes.

In the perspective of the formation of IMBHs proposed in this Letter, the
inflow of gas into the stellar-mass black hole cluster is realized through the accretion of
the surrounding medium onto such a cluster. With enough gas accreted,
the black hole cluster starts to shrink, as in the model of \citet{Kroupa20}.
The timescale on which the shrinking black hole cluster flooded
by gas reaches the relativistic second phase of its collapse, $t_\mathrm{rel}$, depends on its radius, $R$, and mass,
$M_\mathrm{BH}$, and on
the mass of the individual black holes, $m_\mathrm{BH}$, and the mass of the gas, $M_\mathrm{g}$,
within the cluster. For the purpose of the estimates given in this Letter, we assume these
quantities are set to representative values $R=0.1$~pc, $M_\mathrm{BH}=10^4\,M_\odot$,
$m_\mathrm{BH}=10\,M_\odot$, and $M_\mathrm{g}=10^4\,M_\odot$.
As we show in the following
sections, black hole clusters of such properties are expected to have formed
in the CMZ of our Galaxy during its last known starburst period
\citep{Nogueras-Lara20}.

Calculation of the shrinkage timescale for the representative cluster gives
$t_\mathrm{rel}\approx1.9$~Myr (see Appendix~\ref{solutions} for details and a
discussion of wider parameter ranges). Hence, such a black hole cluster
naturally collapses to an IMBH, which suggests a new formation
scenario for IMBHs in the Galactic CMZ.
%
%--------------------------------------------------------------------------------------------------
%
\vspace*{-4mm}
\section{Massive clusters formed during starbursts}
\label{starbursts}
The star formation rate and the properties of the emerging star clusters depend on the
astrophysical conditions in the subject region, such as the matter density or metallicity.
It has been shown \citep[e.g.][]{Weidner04} that in starbursting regions where these
conditions are favourable, the newly born star clusters can be as massive as
$10^{5-7}\,M_\odot$. For the canonical initial stellar mass function \citep{Kroupa01,Kroupa24},
$\mathrm{d}N/\mathrm{d}m\propto{}m^{-2.3}$, such clusters contain
$2.5\times10^{2-4}$ stars initially more massive than $20\,M_\odot$ that end their lives as
black holes. However, combined theoretical and observational investigations suggest that the
initial
mass function in massive star clusters is rather top-heavy \citep[][Fig.~2 therein]{Gjergo25}.
If a top-heavy initial mass function with a power-law index equal to -1 is
considered, the number of stars with initial mass higher than $20\,M_\odot$ increases to
$10^{3-5}$. For the representative mass of each
of the black holes, $m_\mathrm{BH}=10\,M_\odot$, the corresponding mass of the black hole
sub-cluster is thus $M_\mathrm{BH}\gtrsim10^4\,M_\odot$. The radius of such a black hole
sub-cluster can be estimated as $R\approx0.1$~pc according to \citet{Arca-Sedda16}.

According to \citet{Nogueras-Lara20}, a starburst period occurred about 1~Gyr ago in the nuclear
stellar disc, a dense stellar structure within the Galactic CMZ \citep{Launhardt02},
possibly due to a close encounter of the Sagittarius dwarf galaxy with the Galactic centre.
During this starburst event, stars with the total mass of about $4\times10^7\,M_\odot$ were
formed within $\approx100~\mathrm{Myr}$. Even though the properties of the resulting star
clusters are not known, the above-mentioned empirical expectations of \citet{Weidner04} as well
as observations of various starburst galaxies \citep[e.g.][]{Levy24} suggest that these clusters
were likely massive enough to form stellar black hole sub-clusters with masses
$M_\mathrm{BH}\gtrsim10^4\,M_\odot$.
%
%--------------------------------------------------------------------------------------------------
%
\vspace*{-1mm}
\section{Accretion of gas onto the clusters}
\label{accretion}
Star clusters that formed in vast regions rich in gas (such as the CMZ of our
Galaxy) accrete gas for much of their lifetimes \citep{Pflamm-Altenburg09}.
The accretion rate and its impact on the
cluster evolution are determined by the astrophysical properties of the cluster momentary
surroundings, such as the gas density, the stellar (remnant) content of the cluster, and 
by the relative motion of the cluster with respect to the gas reservoir. In full, this
represents a very complex stellar and hydrodynamical problem that includes a plethora of
astrophysical processes. Here, we point out the
main relationships, and leave more detailed investigations for future work.

For the description of the accretion, we adopt the Bondi-Hoyle-Lyttleton model
\citep{Hoyle39,Bondi52} in which the matter accretion rate,
$\mathrm{d}m/\mathrm{d}t$, onto the central object of mass $m$ can be written as
\begin{equation}
\label{bondi}
\frac{\mathrm{d}m}{\mathrm{d}t}\approx4\pi\rho_\mathrm{g}
\frac{\left(Gm\right)^2}{\left(v^2+c_\mathrm{s}^2\right)^{3/2}}~,
\end{equation}
with $\rho_\mathrm{g}$ being the mass density of the gas in the reservoir far away from the
accreting object, $c_\mathrm{s}$ the speed of sound in the undisturbed gas, $v$ the relative
velocity of the object with respect to the gas reservoir, and $G$ the gravitational constant.

Accretion of gas onto a cluster of stars or stellar remnants can be viewed from two
perspectives. In the first, the cluster as a whole is considered to be the accreting body
with the combined mass of its members. The second approach is to
quantify the accretion rates for the individual cluster members
independently and sum these together. It has been shown by \citet{Kaaz19}
that the correct description lies somewhere between these two approaches and
depends on the internal structure of the cluster. In agreement with the intuitive
understanding, denser clusters with more massive members act
more as single-body accretors. Since we primarily focus on dense black hole sub-clusters
from the cores of their parent star clusters stripped by the Galactic tidal
forces, we here adopt the single-body accretor approach.

It is commonly accepted that the gas particle density, $n_\mathrm{g}$, within the
CMZ of our Galaxy is about $10^4~\mathrm{cm}^{-3}$ on average. The actual
value of $n_\mathrm{g}$, however, strongly depends on the particular location as it
reaches up to $10^7~\mathrm{cm}^{-3}$ in the densest parts of the molecular clouds, while it
falls to $10^2~\mathrm{cm}^{-3}$ in regions poorer in gas
\citep[][and references therein]{Mills18}. The average value of $n_\mathrm{g}$ implies
$\rho_\mathrm{g}\approx250\,M_\odot~\mathrm{pc}^{-3}$, which we take as a representative
number for the estimate of the accretion rate (\ref{bondi}) onto the black hole cluster.

Newly born star clusters inherit the motion patterns of their parent molecular clouds
in the Galactic potential and keep these until the combined effect of subsequent encounters
with other massive bodies and possible secular perturbations change them. The velocity
dispersion, $\sigma_\mathrm{cl}$, of the clusters is thus roughly equal to the velocity
dispersion, $\sigma_\mathrm{m}$, of the clouds for some period of time,
$\sigma_\mathrm{cl}\approx\sigma_\mathrm{m}$. The length of this period depends on the
hostility of the cluster environment. For the CMZ and galactocentric
distances of about 100~pc, it is reasonable to
assume that the equality holds for at least a few orbital periods, i.e. a few dozen million years
(see Appendix~\ref{friction} for the dynamical friction inspiral time estimate).
During this period of time, the most massive stars evolve into black holes and segregate
in the cluster centre \citep{PortegiesZwart02}. Simultaneously, the Galactic
tidal stripping causes a loss of the less massive stars from the cluster outskirts,
transforming the cluster into a dark star cluster \citep{Banerjee11,Wu24,RostamiShirazi24}
that consists of the central black hole sub-cluster orbited by a significantly reduced number
of luminous stars with masses lower than a few $M_\odot$.
This is convenient as the absence of massive stars with potentially strong stellar winds
facilitates the gas accretion onto the black hole cluster. The cluster feedback is
driven only by the accretion of some amount of the infallen gas directly onto the individual
black holes (see Appendix~\ref{feedback}).

The molecular clouds in the CMZ of our Galaxy are mostly located in a
twisted ring with a radius of about 100~pc. The overall velocity dispersion of the gas,
$\sigma_\mathrm{g}$, at such galactocentric distances has recently been
determined to be $\sigma_\mathrm{g}\approx30$~km/s \citep[][Fig. 14 therein]{Schultheis21}.
The molecular clouds in the ring, however, appear to orbit
the innermost parts of the Galaxy (including the SMBH Sgr~A$^\star$)
coherently. Their relative motions are thus considerably slower and so
$\sigma_\mathrm{m}<\sigma_\mathrm{g}$. The particular value of $\sigma_\mathrm{m}$ depends
on the details of the dynamics of the clouds in the ring. For the purpose of our
estimate, we assume $\sigma_\mathrm{m}\approx10$~km/s.

The temperature of the gas in the clouds is low. This suggests that the accretion of the gas onto
the clusters is dominated by their motion through the clouds rather than the thermodynamics
of the clouds, $v\gg c_\mathrm{s}$.

For the representative cluster of black holes with mass $M_\mathrm{BH}\approx10^4\,M_\odot$,
$v\approx\sigma_\mathrm{cl}\approx\sigma_\mathrm{m}\approx10$~km/s, and
$\rho_\mathrm{g}\approx250\,M_\odot~\mathrm{pc}^{-3}$, the accretion rate (\ref{bondi})
equals $5.9\times10^3\,M_\odot~\mathrm{Myr}^{-1}$. At this rate, the black hole cluster
accretes gas of the total mass, $M_\mathrm{g}$, equal to its own mass
$M_\mathrm{g}\approx M_\mathrm{BH}\approx10^4\,M_\odot$ in about 2~Myr.
This is fast enough for the assumption of low relative velocity discussed above
to hold through the whole accretion period.
%
%--------------------------------------------------------------------------------------------------
%
\vspace*{-3mm}
\section{Discussion}
\label{discussion}
In this Letter we have formulated a novel IMBH formation scenario with two dynamically
distinct phases. In the first, the cluster of stellar black holes is accreting gas from its
surroundings and shrinking due to the gas drag and mass accumulation as a result. Within a few
million years, the velocity dispersion of the black holes increases enough for the second phase
driven by
the emission of gravitational waves to bring the cluster to its final collapse to the IMBH on
a timescale orders of magnitude shorter \citep[see][]{Kroupa20}.

So far, we have assumed that the black hole cluster parameters are set to their representative
values (see Section~\ref{shrinkage}). Variation of these values affects not only the resulting mass
of the IMBH, $m_\mathrm{IMBH}$, but also the efficiency of the individual processes included
in its formation.

The accretion rate (\ref{bondi}) for the black hole cluster depends quadratically
on its total mass, $M_\mathrm{BH}$. This indicates that clusters with
$M_\mathrm{BH}<10^4\,M_\odot$ are likely to accrete sufficient gas within their lifetimes
only during encounters with the densest parts of the molecular clouds, which are rare. In
contrast, clusters with $M_\mathrm{BH}>10^4\,M_\odot$ accrete the gas very efficiently.
Furthermore, massive black hole clusters originating from massive
star clusters are more compact \citep{Arca-Sedda16}, and thus denser, which makes them
better justified single-body accretors (see Section~\ref{accretion}) in terms of the
quantitative single-body accretion criterion of \citet{Kaaz19}. We note, however, that this
criterion was derived under the assumption of uniform spatial distribution of the cluster
members. In real astrophysical clusters, the member spatial density increases towards their
centres.

The mass function of the black hole cluster also influences its accretion rate in a way
such that a higher abundance of massive black holes facilitates the accretion, while the
opposite is true for low-mass black holes (see Appendix~\ref{feedback}).
In addition, larger accretion radii of the more massive black holes improve the conditions
for the single-body accretor regime for the whole cluster, similarly to the above-mentioned
higher cluster density. For these reasons, top-heavy initial mass functions of the
parent star clusters are favourable for the gas accretion onto the black hole clusters
formed from them.

The shrinkage of the black hole cluster under the assumptions of the model of \citet{Kroupa20}
is similarly efficient for all reasonable values of the cluster parameters (see Fig.~\ref{maps}
in Appendix~\ref{solutions}). Investigating the efficiency of the subsequent relativistic
collapse to the IMBH in the conditions of the still ongoing gas accretion
is a complex problem that requires dedicated efforts that are beyond the scope of this
Letter. However, it has been discussed by \citet{Kroupa20} that the mass-loss due to the
gravitational wave emission can be estimated as $\approx5\%$. The impact of the kicks imparted
during black hole mergers on IMBH formation is likely to be less significant for
higher-mass clusters, due to their deeper gravitational potential well and stronger accretion
gas inflow.

In general, we find that the IMBH formation scenario suggested in this Letter is more
feasible for massive black hole clusters ($\gtrsim10^4\,M_\odot$) originating from massive
star clusters ($\gtrsim10^5\,M_\odot$) with a top-heavy initial mass function, which ensures
a larger number of high-mass black holes. Such clusters can form during starburst periods.

The overall star formation efficiency for such starburst periods is not known.
Even if we assumed, however, that during the starburst in the CMZ
of our Galaxy \citep{Nogueras-Lara20}, much of the gas there had been transformed to stars,
the Galactic bar
would have likely been able to replenish the gas reservoir for the subsequent accretion onto
the black hole clusters within a few dozen million years, given its currently estimated gas
transport rate of 0.1--$1\,M_\odot$~yr$^{-1}$ \citep{Morris96}.
%
%--------------------------------------------------------------------------------------------------
%
\subsection{Observed intermediate-mass black hole candidates}
\label{observed}
Intermediate-mass black holes are long-lived objects, and it can thus be expected that their
orbits in the CMZ evolve on longer timescales due to two-body relaxation
and related effects such as the dynamical friction. For an IMBH of the representative mass
$m_\mathrm{IMBH}=10^4\,M_\odot$ formed in the $\approx100$~pc ring of molecular clouds, the
characteristic time of the inspiral, $t_\mathrm{df}$, owing to the dynamical friction
can be estimated as $t_\mathrm{df}\approx3.8$~Gyr (see Appendix~\ref{friction}).

Such a timescale suggests that IMBHs could have significantly migrated
from the orbits on which they formed not only within the age of the Universe, but also since the
last major starburst in the Galactic centre about 1~Gyr ago \citep{Nogueras-Lara20}.
This is consistent with the fact that all of the five currently observed IMBH candidates
in the CMZ \citep{Maillard04,Oka16,Tsuboi17,Takekawa19a,Takekawa19b,Takekawa20}
are located well within the ring of molecular clouds,
making them compatible with the IMBH formation scenario suggested in this
Letter. We caution, however, that in order to place the predictions of this scenario
into the proper observational context, a more detailed follow-up modelling
of its individual aspects is crucial.

The inspiral due to the dynamical friction can be further complemented by an increase in the
eccentricity of the IMBH orbit thanks to its interaction with the
population of significantly more massive objects, such as the giant molecular clouds
\citep{Perets07}. In such a way, the IMBHs can draw even closer
to the central SMBH Sgr~A$^\star$. This is of special relevance
for one of the IMBH candidates that is found only about 0.14~pc in projection from
Sgr~A$^\star$, within the IRS~13E complex
\citep{Maillard04, Tsuboi17, Peissker24}.
%
%--------------------------------------------------------------------------------------------------
%
\subsection{Further considerations}
On their orbits through the CMZ, the newly formed IMBHs
continue to accrete gas from their momentary surroundings at rates similar to those of
the stellar black hole clusters derived in Section~\ref{accretion}. This accretion can lead to
a direct increase in their mass, but also to the formation of accretion discs around them in which
new generations of stars can form. Such a star formation channel is considered to account
for the presence of the young stars observed in the innermost parsec of our Galaxy on orbits around
Sgr~A$^\star$
\citep[e.g.][]{Levin03}. The stellar black holes originating from these new stellar populations
can eventually coalesce with the central IMBH in the way
described in this Letter, thus causing its possibly significant sequential growth.

A similar recurrent accretion may also occur for initially less massive star clusters with
$M_\mathrm{BH}\approx10^{2-4}\,M_\odot$ and
may lead to a gradual growth of their black hole sub-clusters. At some point, these may become
massive enough to form an IMBH. Such an evolution is possible for the
Arches and Quintuplet clusters that are nowadays observed in the Galactic centre and whose
masses are estimated to be $\gtrsim10^4\,M_\odot$ \citep{Figer99}.
%
%--------------------------------------------------------------------------------------------------
%
\vspace*{-2mm}
\section{Conclusions}
\label{conclusions}
In this Letter we have investigated the accretion of interstellar medium onto clusters of
stellar black holes in order to assess the feasibility of their collapse into intermediate-mass
black holes driven by the gas inflow \citep[as described by][for the case of supermassive
black holes at the centres of forming galaxies]{Kroupa20}. The black hole clusters
are thought to represent the remnants of the stellar dynamical evolution of their parent
star clusters (dark star clusters) formed in gas-rich regions such as galactic central
molecular zones.

Using the Bondi-Hoyle-Lyttleton accretion model, we have estimated the accretion rate for
a $10^4\,M_\odot$ black hole cluster located in the central molecular zone of our Galaxy whose
astrophysical properties are best known. Such massive black hole clusters are likely to have
emerged during the last known starburst period in the Galactic centre reported to have occurred
about $\approx1$~Gyr ago. We have found that,
owing to the high gas densities in the Galactic
central molecular zone and the coherent motions of the molecular clouds located in the twisted
ring with the radius of $\approx100$~pc, the black hole cluster accretes gas of the same mass,
$10^4\,M_\odot$, in about 2~Myr or less. Based on the model of
\citet{Kroupa20}, this creates favourable conditions for the cluster to directly collapse
to an intermediate-mass black hole within additional $\approx2$~Myr.
Even if the ring of molecular clouds is a generally convenient part of the Galactic centre
for intermediate-mass black hole formation, a black hole cluster can undergo its collapse
at any other location if it encounters a dense enough gas cloud at a low enough relative velocity.

We have further shown that the effect of dynamical friction leads to a migration of the newly
formed intermediate-mass black holes towards the Galactic central supermassive black hole
Sgr~A$^\star$ on the timescale of a few billion years. This is compatible with the observed
locations of the reported intermediate-mass black hole candidates found within the distance
of a few tens of parsecs from Sgr~A$^\star$. Such a conclusion holds even if
their formation at the location of the current ring of molecular clouds during the starburst
event $\approx1$~Gyr ago is assumed.

It is likely that the astrophysical conditions in the central molecular zone of our Galaxy are
not exceptional among galaxies of similar type. This suggests that the intermediate-mass black
hole formation mechanism described in this Letter is widely applicable.

Since the inherent
collapse of a stellar black hole cluster into an intermediate-mass black hole is accompanied by
an emission of gravitational waves, a theoretical investigation of their pattern together with an
observational campaign to detect them present a direct way to test the hypothesis formulated
here. Similarly, a more detailed investigation of the radiative signatures of the stellar black
hole cluster collapse in the gaseous medium as well as the preceding accretion process could
possibly provide a link to an already observed class of sources of high-energy radiation or
establish a new class of such sources.
%
%--------------------------------------------------------------------------------------------------
%
\begin{acknowledgements}
We thank the anonymous referee for their useful comments that helped to improve this manuscript.
PK acknowledges support through the DAAD Eastern-European Bonn-Prague exchange programme.
\end{acknowledgements}
%
%--------------------------------------------------------------------------------------------------
%

%
%--------------------------------------------------------------------------------------------------
%
\begin{appendix}
\section{Black hole cluster shrinkage time}
\label{solutions}
\begin{figure}
\centering
\includegraphics[width=\columnwidth]{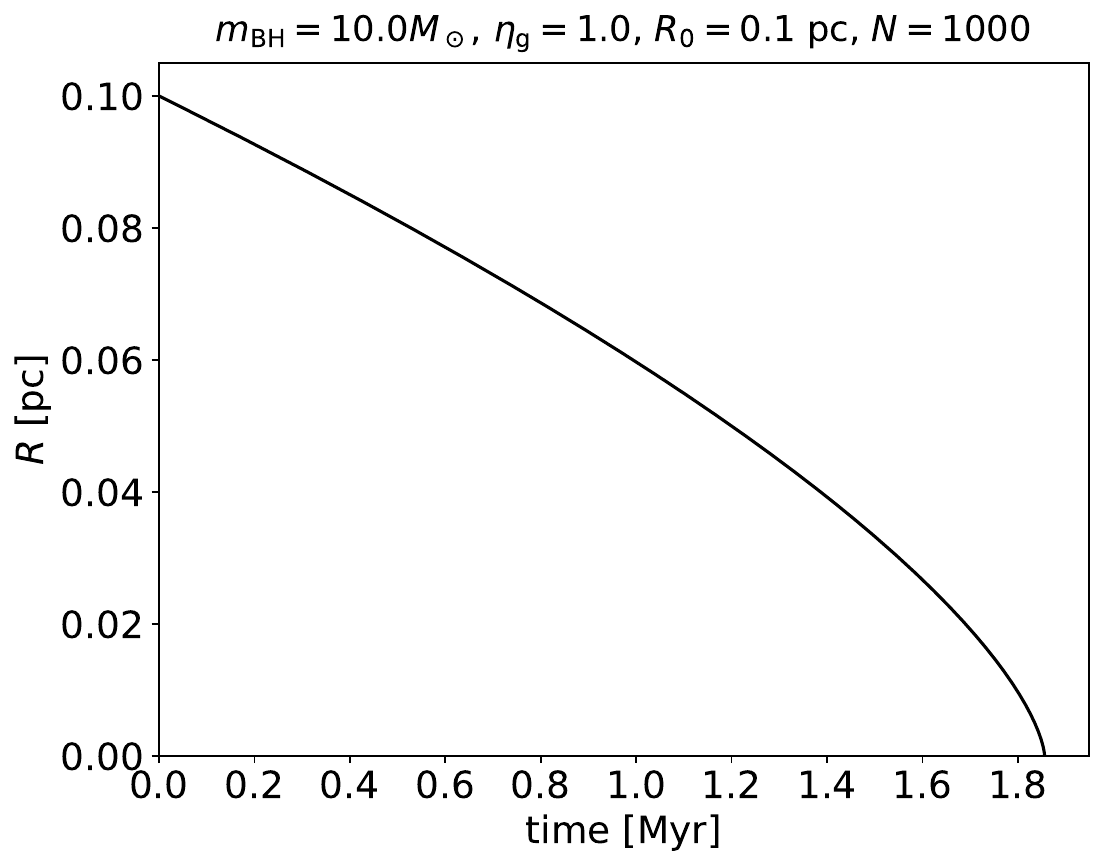}
\caption{Temporal evolution of the black hole cluster radius, $R$, from its representative
initial value $R_0=0.1$~pc. The cluster consists of $N=1000$ equal mass black holes with
$m_\mathrm{BH}=10\,M_\odot$. The total mass of the gas, $M_\mathrm{g}$, within the cluster
is set to $M_\mathrm{g}=\eta_\mathrm{g}M_\mathrm{BH}=\eta_\mathrm{g}Nm_\mathrm{BH}=10^4\,M_\odot$,
i.e. $\eta_\mathrm{g}=1$. The cluster reaches the relativistic phase of its collapse at time
$t_\mathrm{rel}\approx1.9$~Myr.}
\label{collapsing}
\end{figure}

\begin{figure*}
\centering
\includegraphics[width=0.95\columnwidth]{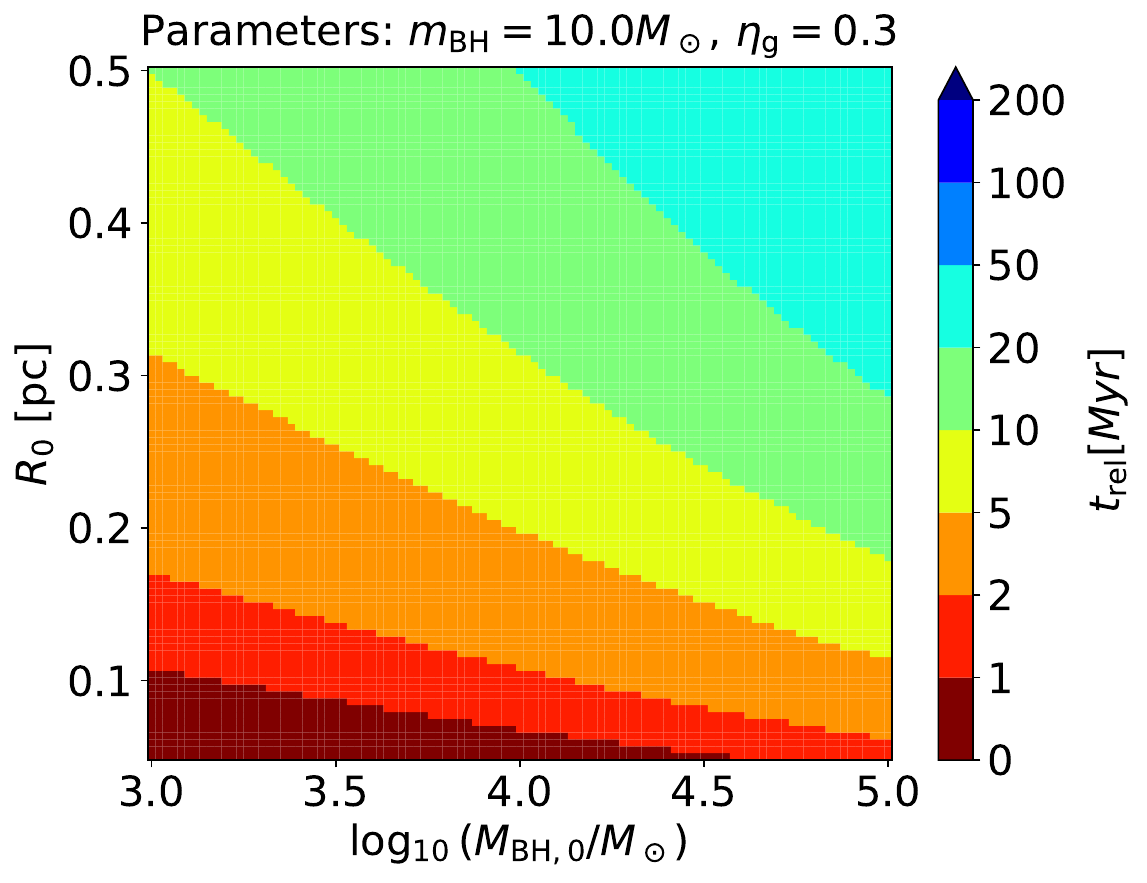}
\hspace{0.5cm}\includegraphics[width=0.95\columnwidth]{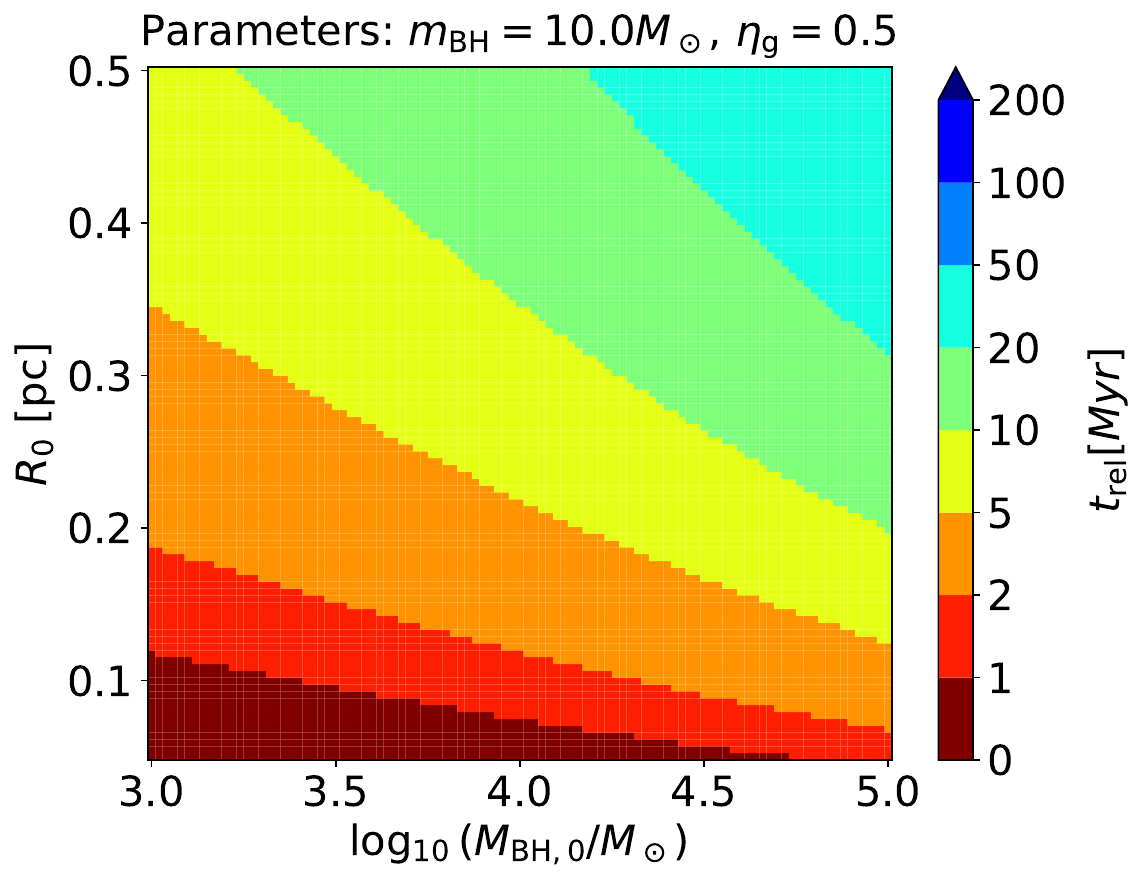}\\[0.5cm]
\includegraphics[width=0.95\columnwidth]{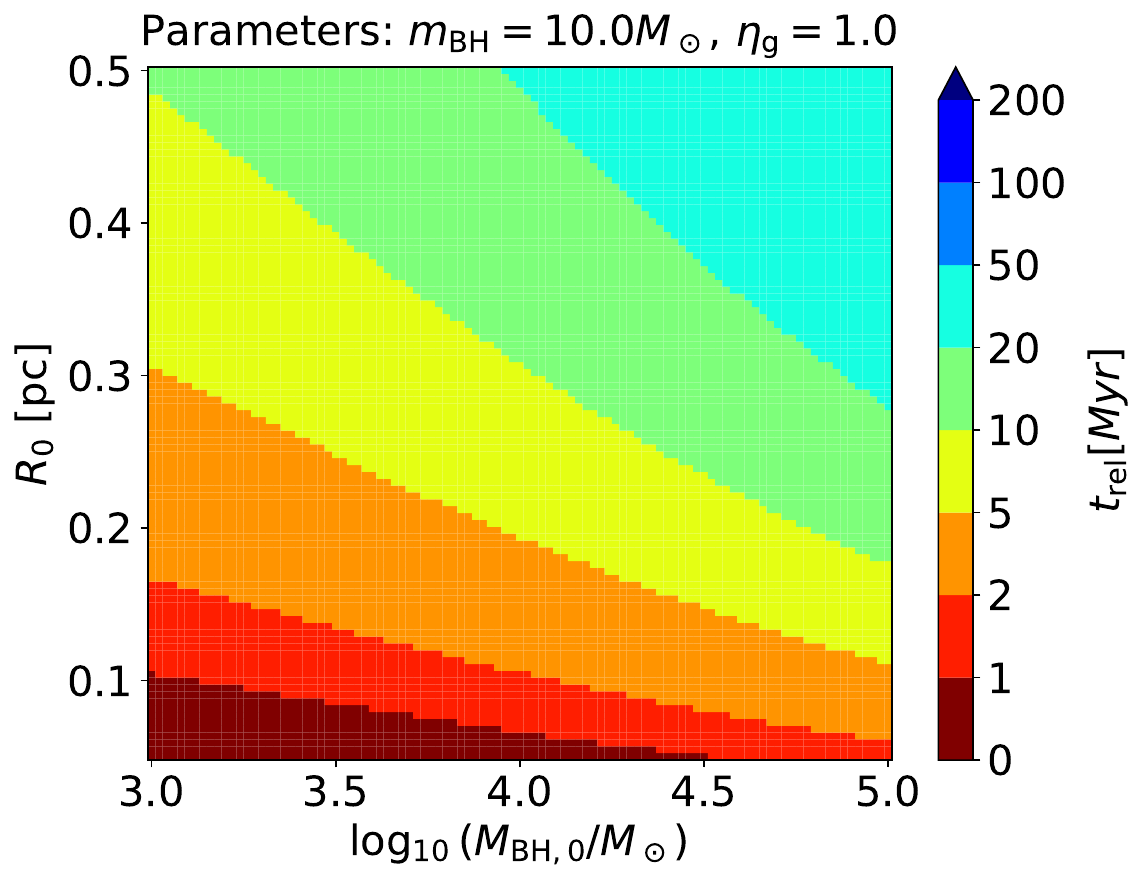}
\hspace{0.5cm}\includegraphics[width=0.95\columnwidth]{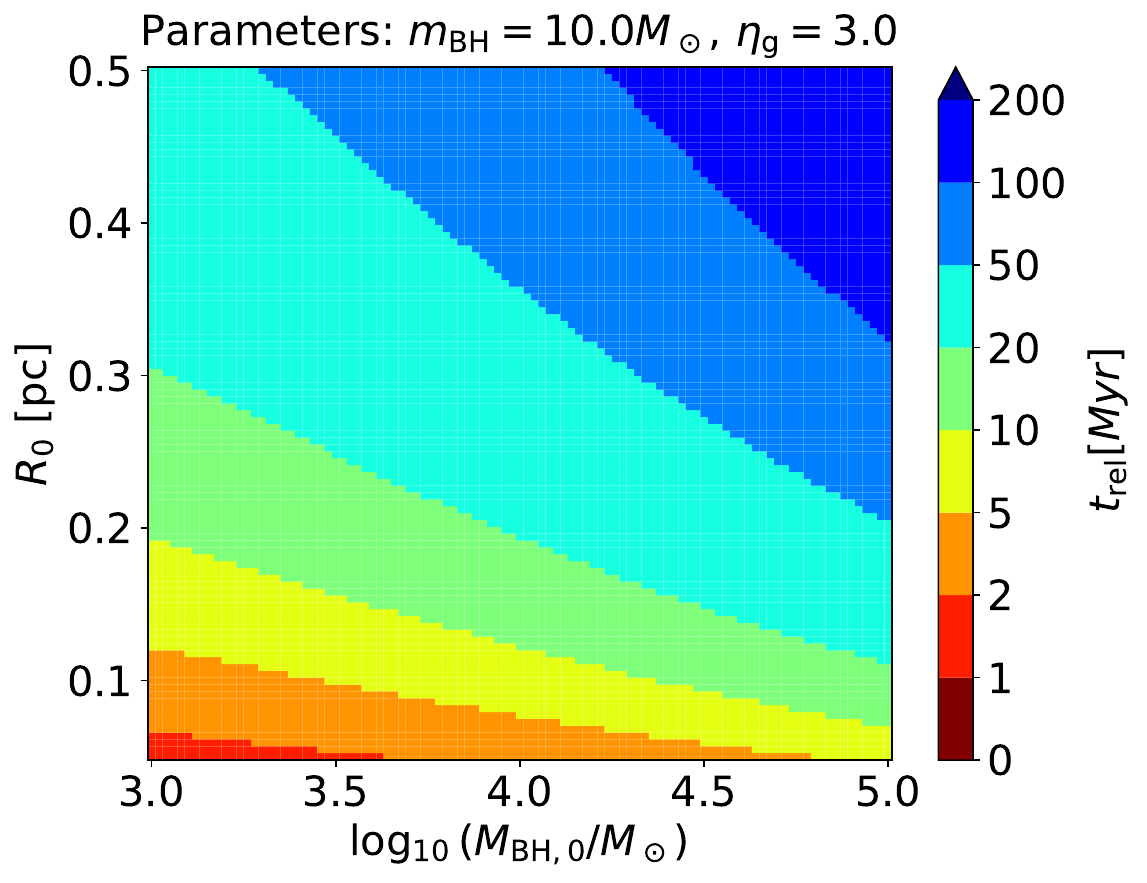}\\[0.5cm]
\includegraphics[width=0.95\columnwidth]{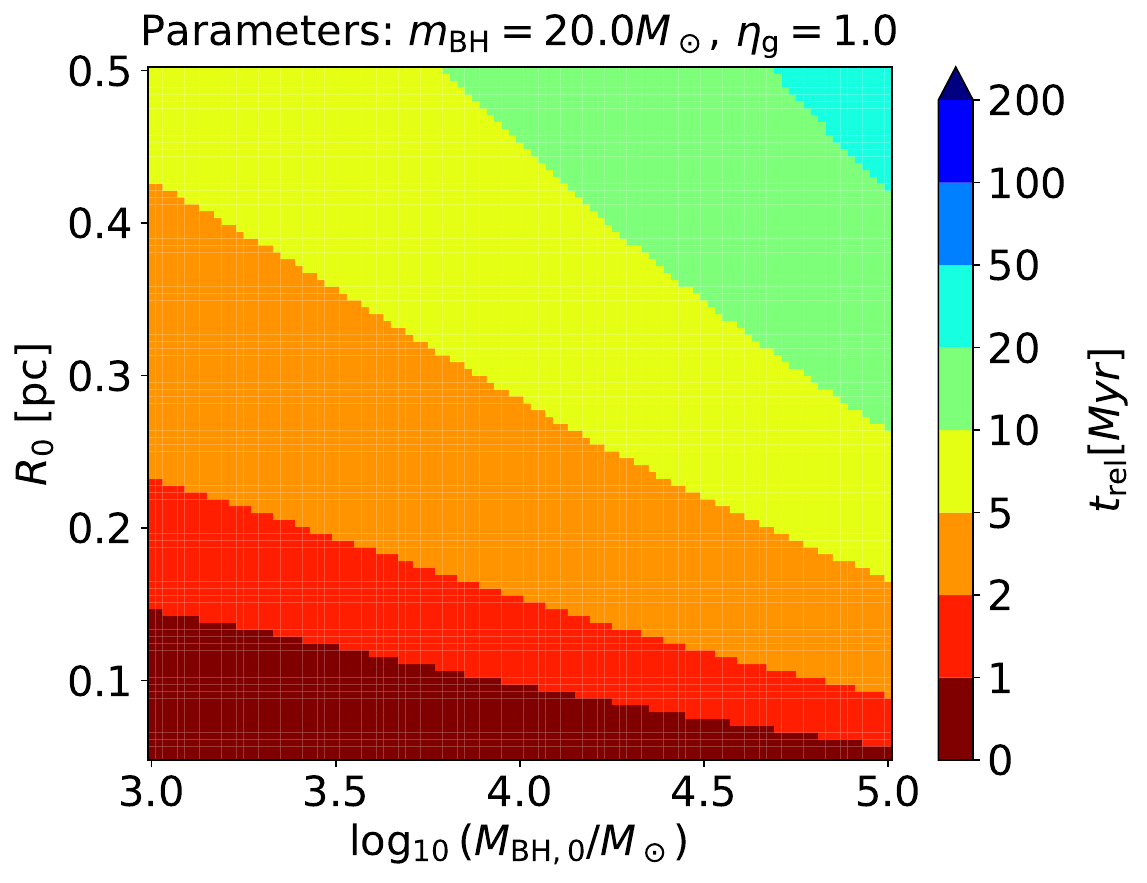}
\hspace{0.5cm}\includegraphics[width=0.95\columnwidth]{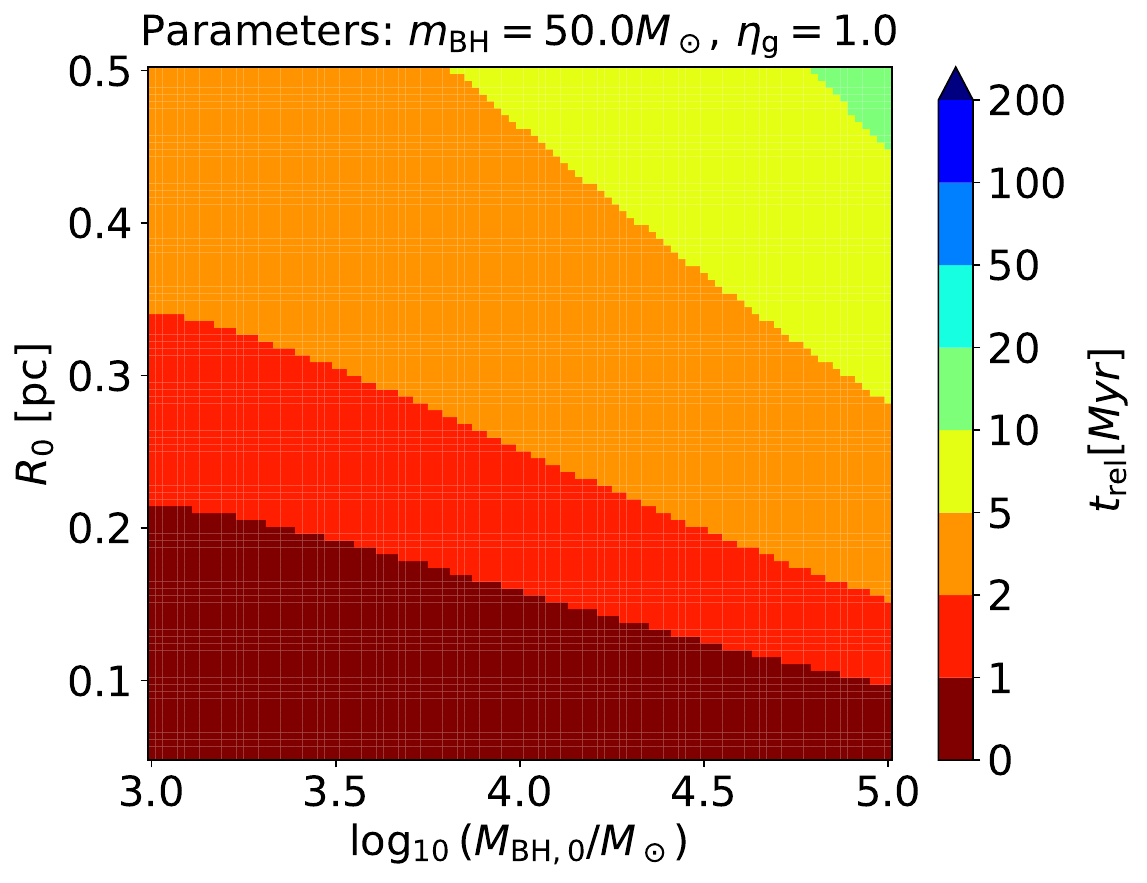}
\caption{Time, $t_\mathrm{rel}$, necessary for the cluster to reach the relativistic phase
of its collapse for various values of its total mass, $M_\mathrm{BH}$
\citep[assumed to be constant and formally denoted by its `initial' value, $M_\mathrm{BH,0}$, for
a more direct comparison with Fig.~8 in][]{Kroupa20}, and initial radius, $R_0$.
The mass of the individual stellar black holes, $m_\mathrm{BH}$, and the
ratio $\eta_\mathrm{g}=M_\mathrm{g}/M_\mathrm{BH}$, where $M_\mathrm{g}$ is the mass of the gas
within the cluster, are given at the top of each panel. Values of the initial radius
$R_0$ cover the interval $\left(0.05,0.5\right)$~pc.}
\label{maps}
\end{figure*}

In this Letter we based our estimates on the stellar-mass black hole cluster with the following
representative properties: $M_\mathrm{BH}\approx10^4\,M_\odot$, $R=0.1$~pc,
$m_\mathrm{BH}=10\,M_\odot$, and $M_\mathrm{g}=\eta_\mathrm{g}M_\mathrm{BH}=10^4\,M_\odot$,
where $\eta_\mathrm{g}$ denotes the ratio of the masses of the gas $M_\mathrm{g}$ and the
black hole cluster $M_\mathrm{BH}$. For such a setting, a
numerical integration of Eq.~(30) from \citet{Kroupa20} that governs the
cluster shrinkage leads to $t_\mathrm{rel}\approx1.9$~Myr (see Fig.~\ref{collapsing}) if
$\mathrm{d}M_\mathrm{g}/\mathrm{d}t=0$ and $\mathrm{d}m_\mathrm{BH}/\mathrm{d}t=0$ is assumed
for simplicity. Resulting values of $t_\mathrm{rel}$ for a wider set of initial settings
are displayed in Fig.~\ref{maps}.
%
%--------------------------------------------------------------------------------------------------
%
\section{Black hole cluster feedback}
\label{feedback}
A fraction of the accreted gas within the cluster inevitably accretes directly onto
the individual black holes, which leads to the generation of high-energy radiation. The
accretion rate, $\mathrm{d}m_\mathrm{BH}/\mathrm{d}t$, for each of the black holes is determined by
the gas mass density, $\rho_\mathrm{gcl}$, within the cluster and the velocity dispersion,
$\sigma_\mathrm{BH}$, of the black holes within the cluster rather than the motion of the
cluster as a whole or the properties of the embedding environment.

The density of the gas within the black hole cluster, $\rho_\mathrm{gcl}$, increases owing to
the ongoing accretion. If we assume that the total gas mass equals the cluster mass,
$M_\mathrm{g}\approx M_\mathrm{BH}\approx10^4\,M_\odot$, and that the characteristic radius,
$R$, of the cluster is about 0.1~pc, we obtain
$\rho_\mathrm{gcl}\approx2.4\times10^6\,M_\odot~\mathrm{pc}^{-3}$, which corresponds to a
gas particle density, $n_\mathrm{gcl}$, within the cluster of
$n_\mathrm{gcl}\approx10^8~\mathrm{cm}^{-3}$.

The velocity dispersion of the black holes, $\sigma_\mathrm{BH}$, can be estimated by the circular
velocity at the characteristic radius $R\approx0.1$~pc and so
$\sigma_\mathrm{BH}\approx\sqrt{GM_\mathrm{BH}/R}\approx21$~km/s. Hence, the accretion
rate (\ref{bondi}) for each of the black holes with mass $m_\mathrm{BH}=10\,M_\odot$
roughly gives $\mathrm{d}m_\mathrm{BH}/\mathrm{d}t\approx5.8\,M_\odot~\mathrm{Myr}^{-1}$.

This is about a factor of 30 higher than the Eddington accretion rate,
$\left(\mathrm{d}m_\mathrm{BH}/\mathrm{d}t\right)^\mathrm{Edd}$, for a $10\,M_\odot$
black hole in pure hydrogen gas under the assumption that the radiation generation
efficiency, $\epsilon$, is set to the typically considered $\epsilon=0.1$. It was shown
by \citet{Johnson22} that for super-Eddington accretion rates that fulfil the condition
$\mathrm{d}m_\mathrm{BH}/\mathrm{d}t
\gtrsim2\left(\mathrm{d}m_\mathrm{BH}/\mathrm{d}t\right)^\mathrm{Edd}/\epsilon\approx
20\left(\mathrm{d}m_\mathrm{BH}/\mathrm{d}t\right)^\mathrm{Edd}$,
the generated radiation is trapped in the infalling gas, allowing the accretion to
proceed unhindered until the gas reservoir is emptied.

The transition into the regime of super-Eddington
accretion with trapped radiation is more
likely for high-mass black holes, due to the fact that the Eddington accretion limit depends
only linearly on the black hole mass, while the accretion rate~(\ref{bondi}) depends quadratically.
Low-mass black holes, on the other hand, are less prone to such a transition.
Which of these two black hole sub-sets in the cluster has a dominant impact on
the gas accretion flow into the cluster depends on the particular shape of its mass function
and on the details of the accretion process itself. In general, however,
the super-Eddington accretion onto the individual black holes in the cluster with trapped
radiation is established once the gas density within the cluster, $\rho_\mathrm{gcl}$, is
sufficiently increased. Before this occurs, the cluster may undergo phases of lower and
higher accretion, depending on the pressure of the radiation generated by the
the momentary accretion flows onto the black holes.

A significantly enhanced accretion takes place in the densest clumps of the giant molecular
clouds where the gas particle density $n_\mathrm{g}\approx10^{6-7}~\mathrm{cm}^{-3}$. In
such clumps, the cluster with parameters set to their representative values (see
Section~\ref{shrinkage}) accretes
$M_\mathrm{g}\approx M_\mathrm{BH}\approx10^4\,M_\odot$ as fast as in
$2\times10^{3-4}$~yr according to Eq.~(\ref{bondi}). With such an accretion rate, the
criterion of \citet{Johnson22} for the transition into the regime with trapped radiation
is immediately fulfilled regardless of the black hole mass.
Given the assumed $v\approx10$~km/s, the cluster covers a distance of about 0.02--0.2~pc
within the cloud over the above period of time, which corresponds to the
size of individual molecular cores \citep[][and references therein]{DeBuizer25}.
%
%--------------------------------------------------------------------------------------------------
%
\section{Dynamical friction time estimate for the intermediate-mass black holes}
\label{friction}
The characteristic time of inspiral owing to the dynamical friction, $t_\mathrm{df}$,
for an IMBH with mass $m_\mathrm{IMBH}$ located initially on a circular
orbit around Sgr~A$^\star$ with radius $r_\mathrm{i}$ can be estimated as
\citep[][Eq. 8.12 therein]{Binney08}
\begin{equation}
\label{inspiral}
t_\mathrm{df}=\frac{38\,\mathrm{Gyr}}{\mathrm{ln}\Lambda}
\left(\frac{r_\mathrm{i}}{100\,\mathrm{pc}}\right)^2\frac{\sigma}{100\,\mathrm{km\,s}^{-1}}
\frac{10^4\,M_\odot}{m_\mathrm{IMBH}}\,,
\end{equation}
where $\sigma$ is the velocity dispersion of the stars and $\mathrm{ln}\,\Lambda$
stands for the Coulomb
logarithm that can be estimated by $\mathrm{ln}\,\Lambda\approx10$
\citep[][Eq. 8.1b therein]{Binney08} for the reference values from equation (\ref{inspiral}).

The typical velocity dispersion of the stars in the surroundings of the IMBH during
its inspiral towards the centre of the Galaxy is higher than within the ring
($\approx100$~pc in radius) of molecular
clouds where it was formed, $\sigma\approx100$~km/s \citep[][Fig. 14 therein]{Schultheis21}.
Hence, for an IMBH with the representative mass
$m_\mathrm{IMBH}=10^4\,M_\odot$ initially on an orbit with radius $r_\mathrm{i}\approx100$~pc,
the inspiral time (\ref{inspiral}) gives $t_\mathrm{df}\approx3.8$~Gyr.
%
%--------------------------------------------------------------------------------------------------
%
\end{appendix}

\end{document}